# A Resource-based Rule Engine for energy savings recommendations in Educational Buildings

Giovanni Cuffaro, Federica Paganelli
CNIT
Florence, Italy
{giovanni.cuffaro, federica.paganelli}@cnit.it

Georgios Mylonas
Computer Technology Institute & Press "Diophantus"
Patras, Greece
mylonasg@cti.gr

*Abstract—* **Raising awareness among young people on the relevance of behaviour change for achieving energy savings is widely considered as a key approach towards long-term and cost-effective energy efficiency policies. The GAIA Project aims to deliver a comprehensive solution for both increasing awareness on energy efficiency and achieving energy savings in school buildings. In this framework, we present a novel rule engine that, leveraging a resource-based graph model encoding relevant application domain knowledge, accesses IoT data for producing energy savings recommendations. The engine supports configurability, extensibility and ease-of-use requirements, to be easily applied and customized to different buildings. The paper introduces the main design and implementation details and presents a set of preliminary performance results.**

## I. INTRODUCTION

Increasing awareness and changing energy consumers' behaviour is considered a key approach towards the achievement of energy consumption reduction [1]. Raising awareness on this topic among young people is considered of primary importance. The Agenda 21 Action plan recognizes the need of increasing sustainable development and energy efficiency education [2]. As a consequence, schools are considered as strategic environments where energy efficiency actions should be deployed. Indeed, in the last decade schools have been the target of studies, education initiatives, as well as energy efficiency actions in several countries [3][4][5].

GAIA [6] is a European research project aiming to tackle energy efficiency in education environments, focusing on behavioural change; in contrast to other similar projects, it does not heavily utilize actuation or other automated approaches. Instead, it uses an IoT infrastructure to monitor buildings in real-time and use such data for customized educational experiences, which aim to engage students in every-day energy-friendly practices. Gamification and a set of educational materials will be used alongside a set of applications to aid teachers and students to achieve better results, in terms of energy consumption in their own school buildings. GAIA solutions will be deployed and validated in more than 15 school buildings (i.e., primary, secondary and high schools) in Greece, Italy and Sweden.

In this work, we present a novel rule engine that, leveraging a resource-based graph model encoding relevant application domain knowledge, accesses IoT data for producing energy savings recommendations. With respect to related work, the adoption of a resource-based model allows the engine to be easily applied in any Web of Things context, and, more generally, to any application context where resources are identified by URIs and exposed through REST APIs. In the paper, we will discuss how design and implementation choices allows the engine to support configurability, extensibility and ease of use requirements, in order to be easily applied and customized to different buildings.

The paper is structured as follows. Section II introduces related work motivating our contribution. In Section III, we present GAIA's main objectives and reference architecture. In Section IV, we describe the resource-based model, design and implementation details of the proposed rule engine. In Section V, we present results of preliminary performance testing activities, while Section VI concludes the paper with insights into our future work.

## II. RELATED WORK

Several works have been proposed to support energy consumption reduction in private and public buildings that address different specific topics. The survey provided by De Paola et al. [7] is an attempt of providing structured and unifying view of main architectures and methodologies for intelligent energy management systems in buildings. The study also surveys possible adoption of sensing technologies and artificial intelligence techniques (e.g., user profiling, user presence detection, activity prediction. etc.). Although most works in the literature focus on enabling automated support for energy savings for supporting and improving the enforcement of energy-savings policies, only a few works focus on providing direct feedback to end users, suggesting actions targeting energy saving. Kobayashi et al. [8] proposes a home energy monitoring system composed by human activity and environment monitoring system and a rule-based energy-saving advice system. Stavropoulos et al. [9] report on the design, development and real-world adoption of an energy saving system in the building of a Greek public university. The system leverages a heterogeneous sensing infrastructure and a semantic middleware on top of which two rule-based approaches for enforcing energy-saving policies are proposed, one based on production rules and the other one using defeasible logics. Mainetti et al. [10] propose a rule-based semantic architecture for enabling users defining and enforcing automation policies in an IoT Context. SESAME-S [11] is as all-in-a-box smart home prototype that uses ontologies for representing the context situation and applies rule-based reasoning through the JESS rule engine [12] for automating the turning on/off of devices and enforcing energy-aware environmental control policies.

While the majority of related work uses a rule-based approach for triggering automation rules, our system aims primarily at producing energy-saving suggestions for the end user, since the objective is to achieve energy efficiency through user behaviour changes. Moreover, our proposed original approach is based on the development of a novel rule engine, designed according to the specific domain's requirements in IoT scenarios. More specifically, rules should be associated to physical and logical entities of the monitored environment, therefore a model of the application domain is needed so rules can be defined on top of it. For instance, a building may be modelled as a tree. Each node may be used to model a physical area inside the building (e.g. a classroom) and deployed sensors and related rules (e.g., anomalous energy consumption during class activities) can be attached to such node. Such representation would allow users to navigate, organize and manage rules in an easy and intuitive way. A further requirement is the need of easily customizing different instances of the a given rule according to the area the rule is attached to. The proposed rule engine allows to specify a rule generic behaviour as a Java class and provide specific settings (e.g. sensor endpoints, threshold values) for each instance.

Several stable rule engines exist (including open source ones), such as Drools and Jess, just to mention some popular and powerful ones. These engines have usually a steep learning curve and it is hard to extend the rule set since rule conditions are expressed in terms of Java object facts. On the contrary, our approach relies on a resource-based and REST-compliant model, which represents sensors, smart things and relevant domain entities and rules as web-addressable resources. REST [13] is considered the main reference architectural style for Web representation and management of real-world objects [14]. The adoption of this resource model in the rule engine design allows an easy way of adding new rule instances or new types of rules at runtime, since conditions are expressed in terms of facts related to REST-based resources.

III. BEHAVIOUR-BASED ENERGY SAVING: GAIA PROJECT

In this section, we provide a brief overview of the goals and overall architecture of the GAIA Project, in order to briefly describe the context motivating the design and development of the proposed rule-based engine for energy saving recommendations.

The Green Awareness in Action (GAIA) EU Project [6] aims at producing a framework comprising Internet of Things deployments, ICT services and applications, as well as education content and actions for increasing sustainable energy awareness among students, while also achieving energy savings through behaviour changes in school buildings. The main objectives of GAIA are listed hereafter:

- *Increasing awareness and engagement in energy efficiency*. GAIA will raise awareness on young people on how the individual behaviour can influence energy consumption at home, at school, etc. Students are the primary target, but also teachers, building managers and administrative and technical staff will be involved.
- *Education*. GAIA will support teachers in the educational process with courses and workshops aiming at increasing students' knowledge and skills on energy efficiency topics.
- *Feedback on energy consumption and behaviour-based recommendations*. Distributed metering and sensing technologies will be used for acquiring data on energy consumption and conditions in school buildings (e.g. activity, occupancy, external and internal weather information). These data will be used to provide direct feedback to people about the building energy consumption profile and analysed for identifying possible actions and recommendations for behaviour-based energy savings. Moreover, GAIA's educational material will be combined with such data, in order to deliver educational activities customized to the context of each school participating in the project.
- *Playing to learn*. GAIA will develop an educational game, accessible through web browsers and smartphones to increase students' awareness and involvement in energy efficient behavioural changes. As with the educational material, real-life data will also be integrated with the gaming activities.

In this paper, we will focus on the contribution towards feedback on energy consumption and behaviour-based recommendations.

The GAIA system includes three main layers (Fig. 1):
- *IoT Infrastructure*, which includes the sensing and power metering infrastructure deployed at each school building to gather monitoring data. Different sensing infrastructures have been deployed in the involved schools.
- T*he GAIA Service Platform*, which is the backend system that provides capabilities for storing and analysing data regarding the energy consumption of buildings and users' behaviour and produces feedbacks to be delivered to end users through applications.
- The *Applications* layer, which handles the interaction with end-users. A *Serious Game* will target students, while a *Web Application* will be developed to provide information and recommendations to building managers. Moreover, in-building *Pervasive Displays* will provide direct access to energy consumption data and savings achievements on a continuous basis.

The *GAIA Service Platform* should acquire data from multiple geographical sites and multiple technological platforms (i.e., heterogeneous *IoT infrastructures*) through heterogeneous APIs and data formats. The *GAIA Service Platform* acts as an intermediate layer that stores measurements acquired from IoT deployments and makes them accessible to end-users' applications through a uniform REST interface, thus hiding differences of the underlying technological infrastructure. This allows applications to have a unified global view of available data, in terms of a unified convention for naming resources and a unified mechanism for accessing them.

The *Recommendation Module* is part of the *GAIA Service Platform*. It acquires information on the energy consumption profile and context of the school (sensor measurements, as well as analytics provided by other GAIA Service Platform modules) and produces energy-saving recommendations, which are then appropriately delivered to end-users through applications. Recommendations provide primarily suggestions for energy-saving behaviours changes (targeting building managers, students and teachers as well), but can also provide building managers with suggestions for technical maintenance interventions or building renewal actions. Produced recommendations should additionally be customizable according to the school building's characteristics (geographical location, internal space allocation, user habits, etc.).

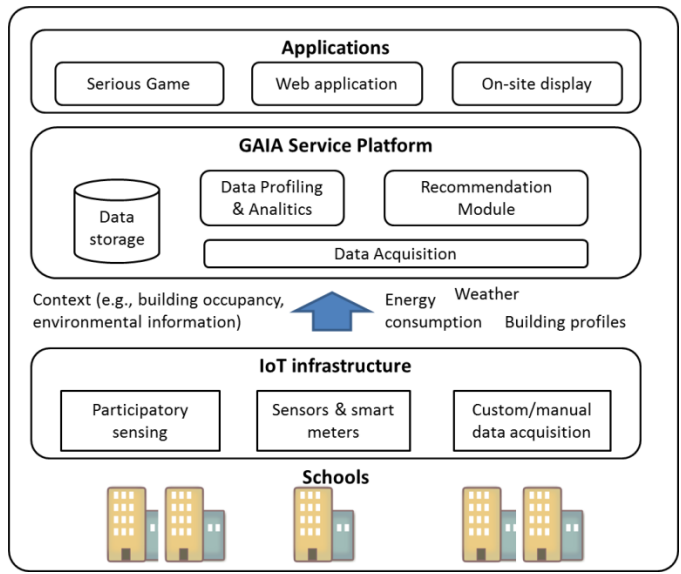

Fig. 1. GAIA conceptual architecture

IV. RESOURCE-ORIENTED RULE-BASED ENGINE

The *Recommendation Module* provides energy saving recommendations in the form of real-time notifications or events to be logged and collected for later use (e.g., report generation). These outputs can be delivered to end users through applications in the form of push notifications or reports produced periodically. The module is implemented as a rule engine which analyses data gathered by the sensor infrastructure and triggers appropriate actions (i.e., delivery of notification messages including the appropriate recommendation and event logging) when the conditions specified by the rules are verified. The rule engine has been conceived in order to meet the requirements of the experimentation activities to be performed in the GAIA project, introduced hereafter:

*Extensibility*: new type of rules can be created from scratch (programmatically) or by editing the configuration of already implemented customizable rules (e.g., specifying an expression to be evaluated) to cope with unforeseen application needs and/or context changes (e.g., new kind of sensor added, new recommendation scenarios, etc.).

*Configurability:* the same type of rule can be applied to create distinct instances to be applied to different schools and different areas of the same school, with appropriate customization of thresholds, sensors' identifiers (i.e., URIs) and content of notification messages.

*Ease of use:* the user should be capable to easily navigate the set of rules associated to a school and to its constituent areas (e.g., classrooms, laboratories, etc.) and modify them as needed (e.g., by adding, modifying and deleting rules).

*A. Resource-based graph model*

The rule engine knowledge of the application domain is represented through a resource-based model (see Fig. 2). A *Resource R* is any entity that can be identified through a URI and can be accessed and manipulated through CRUD operations via HTTP protocol (in accordance to main principles and best practices of the REST architectural style. We define the following main types of Resources: *Area*, *Sensor*, *Parameter* and *Rule*.

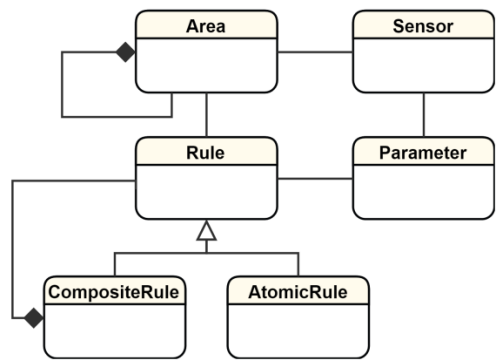

Fig. 2. Resource-based model

The *Area* represents a physical or symbolic location/area, which can be further characterized with information pertaining the domain of interest (e.g., an Area representing a school building can be enhanced with information regarding number of students, surface, yearly energy consumption, etc.). A *Sensor* represents a physical sensor that gathers monitoring parameters (i.e., resources of *Parameter* type) covering a specific *Area*. A *Parameter* may represent a physical parameter (e.g., temperature) gathered by a *Sensor* or a parameter derived from physical ones (e.g., relative humidity). A *Rule* is a resource whose internal behaviour consists in verifying a condition and, if the condition holds, triggering an action. A Rule is further specialized as an *AtomicRule* or a *CompositeRule*. An *AtomicRule* is a rule whose condition is specified in a self-contained way, while a *CompositeRule* is defined as a composition of conditions of other rules (children rules), which can be *AtomicRule* as well as *CompositeRule*. We have the following types of composition:
- *AnyCompositeRule*: it is triggered if any of the children's conditions are true (logical OR behaviour).

- *AllCompositeRule*: it is triggered if all the children's conditions are true (logical AND behaviour).
- *RepeatingRule*: it verifies if in a given time interval, the condition of a child rule has been verified more than a configurable number of times.

Fig. 3 provides an example of the resource-based graph generated when instantiating the above-mentioned model in a specific context (i.e., a school involved in GAIA experimentation activities). The root node represents the *School Area*, which contains other *Areas*. In our example the school building is made of a set of *Areas*: a *SportBlock* (e.g., the gymnasium hall) and a *TeachingBlock*, containing *Classrooms*, *Laboratories* and a *Hall*. A *PowerFactor Rule*, checking if the power factor is below a given threshold, is assigned to the *TeachingBlock* and the *Hall Areas*, using as input the *PowerFactor Parameter* measured by the corresponding *Sensors*. A *Rule* that evaluates the level of external luminosity and the use of artificial light (by analysing the value of absorbed *Active Power*) to suggest energy saving actions (e.g., switch off the light and exploit the natural light) is assigned to the *Hall*. A *ComfortIndex Rule* is assigned to the *Sport Block* to evaluate the heat index provided by the joint evaluation of *Temperature* and *Relative Humidity*, measured by *Sensors*.

## B. Implementation

The rule engine has been implemented as a Java application, using the Spring Framework. The engine is responsible for executing user-defined rules, according to the resource-based model described above. The resource-based graph that encodes the application domain knowledge is persisted in an external database and loaded at runtime by the rule-engine to instantiate the corresponding rule objects. More precisely, at runtime the resource-based graph model is traversed and, for each encountered rule, the structure of the rule is replicated in the engine, which instantiates the rule class, characterizing the actual instance filling the class attributes with the values retrieved from the DB, leveraging Java Reflection API.

### 1) Rule engine logic

The abstract class *GaiaRule* defines the basic structure and behaviour of a rule in the engine logic. It provides common attributes, default implementation of basic methods and some services implementing the I/O behaviour. The *GaiaRule* attributes are listed hereafter: *name* (the name of the rule), *description* (a brief description of the rule), *suggestion* (a brief textual recommendation suggesting energy saving actions and behaviour change), the *URI* (the identifier of the Rule), *Parameters URIs* (the endpoints to gather measurement values through REST APIs).

All rule instances are a subclass of the *GaiaRule* abstract class, which defines the general behaviour of a rule through the following methods:

- *init()*: executes the needed initialization tasks, if needed, and validates the rule (i.e. it checks if all the required fields are loaded). The result of this method is checked when the rule is loaded, if false the rule is discarded.
- *condition()*: verifies if the condition is met, this is left as an abstract method, that has to be implemented in a subclass.
- *action()*: the default implementation consists in sending a basic push notification and logging the related event, leveraging the WebSocket and the EventLogger services, respectively, described below.
- *fire()*: it invokes the *condition()* method and, if the condition is verified, invokes the corresponding *action()*.

The rule engine also implements some capabilities that can be re-used by rule instances: a WebSocket service, which allows to deliver a message through a WebSocket channel to interested applications; a Measurements service, which fetches the needed measurements from the Parameters URIs only once at the beginning of each iteration; a Database service, which provides access to the persistence layer; an EventLogger service, which, when a rule is fired, logs the most significant properties of the occurred event.

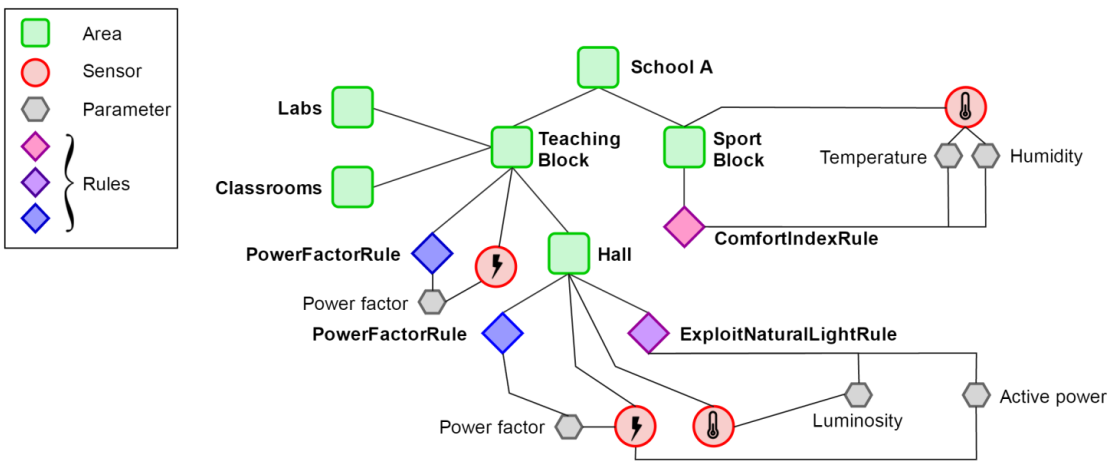

Fig. 3. Example of a Resource-based graph representing energy-related entities, Areas, Sensors, Parameters and Rules for a given school (School A)

The above-described rule engine logic leverages the resource information model and handles corresponding *GaiaRule* instances in order to guarantee the user with various levels of configurability and extensibility. With the term user, we refer here to the developer, accessing the source code and a GUI, and an appropriately trained business user (i.e., building manager), capable of accessing a GUI for managing rules.

Indeed, the rule engine logic allows to define the following types of *GaiaRule*-derived concrete classes: Custom Rules and Template Rules. Custom Rules are direct subclasses of the GaiaRule, whose behaviour in terms of *condition()* and *action()* methods implementation, is defined programmatically. The value of some configurable parameters can be customized for each different instance of a Custom Rule (e.g. threshold and Parameters URIs values, as shown in Fig. 4). Template Rules define only a part of their behaviour, since its specification is finalised through a set of configurable parameters and fields. We provide some examples for the sake of clarity. For instance, a template rule class provides a basic behaviour for checking a condition expressed as comparison of a measurement value with a threshold. The type of comparison operator (e.g. less than, greater than) can be specified as any other parameter (i.e. loaded from the database) when instantiating the rule (see Fig. 4). A second, more general, example is a template rule class that checks the condition by verifying an expression provided as input parameter, together with the Parameter names and URIs (e.g., a formula combining the values of given Parameters).

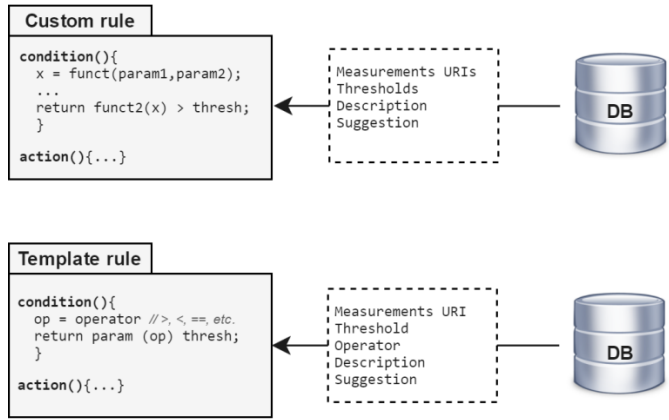

Fig. 4. Examples of a Custom Rule (top) and a Template Rule (bottom)

*2) Persistence Layer*

The persistence layer is realized with OrientDB 2.2.x., a hybrid Graph/Document NoSQL database. We chose this software because it offers a persistence model that accommodates well our resource graph-based model. Moreover, the adopted database system offers a set of REST APIs for easy access to the resources and comes with an intuitive web GUI that can be used to create and edit the resource information model, including the structure and the parameters of the rules.

In our implementation, we have created a class in the OrientDB schema for every Java class representing a rule (with the convention of using the same name for the sake of clarity); each class in the DB should contain the configurable fields needed by the rule to be executed. Since the schema supports class inheritance every rule inherits some basic properties from the *GaiaRule* reflecting the application model. Table I provides an example of the fields of a *ComfortIndex Rule* instance stored in the DB. Once the schema has been created it is possible to define the rules instances in the database, fill the required properties and link rules and areas into a tree structure.

As already mentioned above, the rule engine application loads this structure traversing the graph persisted on the database, instantiates the right classes at runtime and fills the required fields using Java Reflection. OrientDB supports both schema-less and schema-full mode, we use a schema-hybrid approach to have classes with predefined properties (some of them mandatory) which can be extended with new fields at instance level. This feature is needed to support the specification of Template Rules.

TABLE I. EXAMPLE OF A RULE INSTANCE STORED IN THE DB

| Rule instance | Comfort index | |
|---|---|---|
| Field name | Description | example |
| @rid | The ID of the vertex in the database | #25:241 |
| @class | The class of the vertex in the database / Java class name | ComfortIndex |
| name | Name of the rule | CI Room 3 |
| description | Description of the rule | [...] |
| suggestion | Brief suggestion to give when triggered | Open the window |
| temperature_uri | URI of the temperature sensor | http://gaia-x/gw1/temp |
| humidity_uri | URI of the humidity sensor | http://gaia-x/gw1/humid |
| threshold | Threshold for the computed comfort index | 32 |

*3) Execution flow*

In the initialization phase, the engine loads the resource model from the database and internally replicate the structure following the composite pattern. This is done by instantiating the right Java classes using the @class property of each vertex stored in the database. For each Rule resource retrieved from the DB, the corresponding Java object is created by filling the object attributes with the values stored in the DB. The object is initialized and a basic validation of the fields is executed, the object is added to the tree only if it is valid.

In the operational phase, the following tasks are performed periodically: the measurement values required by the rules are acquired through appropriate requests to the Parameters URIs at the beginning of each iteration and shared among all rule instances, the *fire()* method is invoked recursively on the composite structure of rules. When a condition of a rule is verified, the corresponding action is performed. The activation of each rule can be scheduled at the desired frequency. Of course, the designer should take care in specifying the appropriate frequency for rule condition verification with respect to the sampling frequency of sensors whose measurements are evaluated in the condition (i.e. the period between two consecutive condition verifications should be greater than the sensor's sampling period).

## V. PRELIMINARY EVALUATION

In this section we report the results of preliminary performance evaluation tests. The testing activities have been carried out to provide an estimation of the time needed by the Java rule engine for instantiating the rule tree loaded from the database. Tests were executed on an Intel Core i5-3340 desktop PC (using 2 cores) with 4GB RAM, running Ubuntu 16.04.1 LTS and hosting the rule engine and the database.

We evaluated the following two metrics: *total_time*, the time needed for loading the resource model from the DB and for instantiating the rules according to the loaded model, and the *instantiation_time*, which is the time needed just for rule instantiation (measured by leveraging DB caching mechanisms). Test iterations have been run by varying the total number of rules (100, 200, 400, 800, 1600, 3200, 6400 and 12800 rules). The input resource-based graphs composed by *Rule* and *Area* vertices have been randomly generated. Each *Area* vertex is recursively associated to 3 children *Areas* and 5 *Rules*, until the target total number of rules is achieved. We used two types of test rules: a simple rule, triggered with a random probability, and a composite rule (*RepeatingRule*) associated with a simple rule (the pair is counted only once in the total number of rules).

Fig. 5 shows the average results over 100 iterations for the *total_time* and the *instantiation_time*. Results show that these latency metrics have a sublinear trend against the total number of rules, which is an encouraging result.

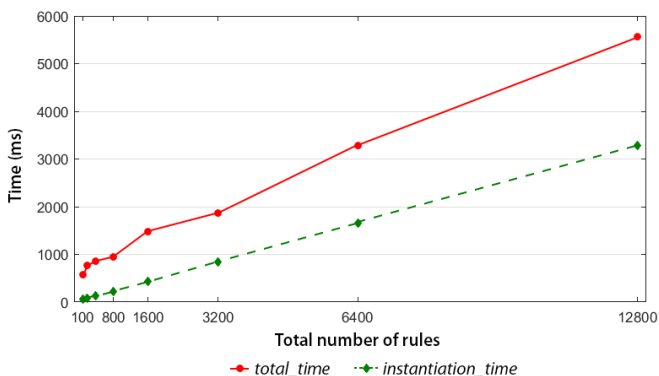

Fig. 5. Time needed for loading the resource model and instantiating the rules (*total_time*) and for mainly instantiating the rules (*instantiation_time*).

## VI. CONCLUSIONS

In this paper, we introduced a rule engine, designed in the framework of energy-efficiency application scenarios of GAIA, aiming at producing energy saving recommendations using IoT data. Nonetheless, the rule engine is general-purpose and can be applied in Web of Things contexts as well as in more general contexts, where resources are identified by URIs and exposed through RESTful APIs. In the near future, the rule engine will be used in GAIA experimentation activities in real-world school scenarios. This activity will help us in effectively validating configurability, extensibility and ease-of-use properties of the rule engine with real users. We will develop a more user-friendly GUI for managing rules and perform additional performance tests in realistic scenarios.

ACKNOWLEDGMENT

This work is supported by the EU research project GAIA, under the Horizon 2020 Programme and contract number 696029. Authors would like to acknowledge the collaboration with the rest of the GAIA consortium. The authors acknowledge the technical support by Mr. Luca Capannesi.